\documentclass[prd,preprint,nofootinbib]{revtex4}

\usepackage{graphicx}
\usepackage{amssymb}
\usepackage{amsmath}
\usepackage{color}

\begin{document}
 
\newcommand{\be}{\begin{equation}}
\newcommand{\ee}{\end{equation}}
\newcommand{\aprime}{\mathbf{a}^{\prime}}
\newcommand{\bprime}{\mathbf{b}^{\prime}}
\newcommand{\kh}{\hat{k}}
\newcommand{\Ip}{\vec{I}_+}
\newcommand{\Imi}{\vec{I}_-}
\newcommand{\bc}{\begin{cases}}
\newcommand{\ec}{\end{cases}}
\newcommand{\cD}{\mathcal{D}}
\newcommand{\xv}{\mathbf{x}}
\newcommand{\qv}{\mathbf{q}}
\newcommand{\pv}{\mathbf{p}}
\newcommand{\trec}{t_{\mathrm{rec}}}
\newcommand{\trei}{t_{\mathrm{rei}}}

\newcommand{\red}{\color{red}}
\newcommand{\cyan}{\color{cyan}}
\newcommand{\blue}{\color{blue}}
\newcommand{\magenta}{\color{magenta}}
\newcommand{\yellow}{\color{yellow}}
\newcommand{\green}{\color{green}}
\newcommand{\rem}[1]{{\bf\blue #1}}

\def\lrfp#1#2#3{ \left(\frac{#1}{#2} \right)^{#3}}

\begin{flushright}
UT-12-06\\
TU-901\\
IPMU12-0027
\end{flushright}


\title{
Gravity mediation without a Polonyi problem
}


\author{Kazunori Nakayama$^{(a,c)}$,
Fuminobu Takahashi$^{(b,c)}$ and
Tsutomu T. Yanagida$^{(c,a)}$
}

\affiliation{%
$^a$Department of Physics,
     University of Tokyo, Tokyo 113-0033, Japan\\
$^b$Department of Physics, Tohoku University, Sendai 980-8578, Japan\\
$^c$Institute for the Physics and Mathematics of the Universe, 
     University of Tokyo, Kashiwa, Chiba 277-8568, Japan
}

\date{\today}

\vskip 1.0cm

\begin{abstract}
Recent indications of the 125\,GeV Higgs at the LHC can be explained 
in a relatively high-scale SUSY scenario where the sparticle masses are multi-TeV
as is realized in the focus-point region. 
However, it suffers from the notorious cosmological Polonyi problem.
We argue that the Polonyi problem is solved and thermal or non-thermal leptogenesis scenario works successfully,
if a certain Polonyi coupling to the inflaton is enhanced by a factor of 10-100.

\end{abstract}


 \maketitle

\section{Introduction}

Recently the ATLAS~\cite{ATLAS-CONF-2011-163} and CMS~\cite{HIG-11-032} collaborations
reported event excesses, which may imply the Higgs boson with mass of about $125$\,GeV.
While it is difficult to explain the Higgs mass  in 
the minimal supersymmetric standard model (MSSM) as long as the 
sparticle masses are around 1\,TeV~\cite{Okada:1990vk},
such a Higgs mass can be explained if the sparticles are heavier than multi-TeV~\cite{Okada:1990gg,Giudice:2011cg,Ibe:2011aa,Feng:2011aa}.
One of such scenarios is the anomaly-mediated SUSY breaking (AMSB) model~\cite{Giudice:1998xp},
where the sfermions and the gravitino are $\mathcal O(100$--$1000)$\,TeV and the gaugino masses are
$\mathcal O(100$--$1000)$\,GeV, given by the AMSB relation.
Phenomenological aspects of this scenario have been discussed in Refs.~\cite{Ibe:2011aa,Moroi:2011ab,Ibe:2012hu},
and it was shown that it is compatible with thermal leptogenesis~\cite{Fukugita:1986hr,Buchmuller:2003gz},
which requires the reheating temperature as high as $T_{\rm R}\gtrsim 10^9$\,GeV~\cite{Giudice:2003jh}.\footnote{
	In this letter $T_{\rm R}$ is defined as $T_{\rm R}\equiv (10/\pi^2g_*)^{1/4}\sqrt{\Gamma_\phi M_P}$
	where $\Gamma_\phi$ is the inflaton decay rate.
	Note that this definition of $T_{\rm R}$ is smaller than that of Ref.~\cite{Giudice:2003jh}
	by a factor $\sqrt{3}$.
}

Another attractive scenario is that all sparticles are $\mathcal O(10)$\,TeV in the gravity-mediated SUSY breaking.
The scenario alleviates the SUSY flavor/CP problems because of the heavy SUSY particles,
while it explains the 125\,GeV Higgs boson for $\tan\beta \gtrsim 5$~\cite{Giudice:2011cg}
and the present dark matter abundance 
(see Ref.~\cite{Feng:2011aa} for realization in the focus-point region~\cite{Feng:1999mn}).
However, the scenario  suffers from the cosmological Polonyi problem~\cite{Coughlan:1983ci,Banks:1993en},
since there must be a singlet SUSY breaking  field, called the Polonyi field, in order to generate
sizable gaugino masses.
Although the Polonyi may decay before the big-bang nucleosynthesis (BBN) begins 
for the Polonyi mass $m_z \gtrsim \mathcal O(10)$\,TeV, it releases a huge amount of entropy
because it dominates the Universe before the decay.
Thus the leptogenesis scenario does not work in this setup.

An interesting solution to the Polonyi problem was proposed long ago by 
Linde~\cite{Linde:1996cx}. It was pointed out that, if the Polonyi field has a large
Hubble-induced mass, it follows a time-dependent potential minimum 
adiabatically and the resultant amplitude of coherent oscillations is exponentially 
suppressed. Recently, two of the present  authors (FT and TTY) noticed that there 
is an upper bound on the reheating temperature 
for the adiabatic solution to work~\cite{Takahashi:2011as} and also showed that such a large Hubble mass may be a consequence of the strong dynamics at the Planck scale~\cite{Takahashi:2010uw}
 or the fundamental cut-off scale one order of magnitude lower than the 
 Planck scale~\cite{Takahashi:2011as}. More important, the present authors found that there are generally additional
contributions to the Polonyi abundance which depends on the inflation energy scale, and 
we  showed that the Polonyi problem is still solved or greatly relaxed in  high-scale inflation 
models~\cite{Nakayama:2011wqa,Nakayama:2011zy}. 
In this solution, we do not need any additional mechanism to dilute the Polonyi abundance.
Therefore, it may revive the conventional Polonyi model as a realistic SUSY breaking model,
which is fully compatible with the current experiments and observations, including the 125\,GeV Higgs boson.

In this letter we study the adiabatic solution in detail, considering various production processes of the 
Polonyi field as well as the thermal and non-thermal gravitino production. In particular, we focus on a minimal model
in which only a certain coupling of the inflaton to the Polonyi field is enhanced. We also consider explicit inflation models
to see if there is an allowed parameter space where the Polonyi and gravitino problems are solved.

\section{The Polonyi model for gravity-mediation}

First we briefly review the cosmological Polonyi problem in the gravity mediation.
Let us denote the Polonyi field by $z$, which makes a dominant contribution to the SUSY breaking.
Its $F$-term is given by
$F_z = \sqrt{3}m_{3/2}M_P$ where $m_{3/2}$ is the gravitino mass and $M_P$ is the reduced Planck scale.
It generally couples to the MSSM superfields as
\begin{equation}
	\mathcal L = \int d^4\theta \left( -c_Q^2 \frac{|z|^2|Q|^2}{M_P^2} \right )
	+ \left( \int d^2\theta c_g \frac{z}{4M_P}W_aW^a+ {\rm h.c.} \right) ,
	\label{KW}
\end{equation}
where $Q$ and $W^a$ collectively denote the matter and gauge superfields, respectively,
and $c_Q$ and $c_g$ are constants of order unity.
Here and hereafter, $c_Q$ and $c_g$ are taken to be real and positive, for simplicity.
These couplings give masses of order $m_{3/2}$ to the SUSY particles, as
\begin{equation}
	m_{\tilde Q}^2 = (c_Q^2+1)m_{3/2}^2,~~~m_{\tilde g} = \frac{\sqrt{3}c_g}{2}m_{3/2}.
\end{equation}
Note that $z$ must be a singlet field in order to give a sizable mass to the gauginos.
The following term in the K\"ahler potential yields the sizable $\mu$ and $B$ terms~\cite{Giudice:1988yz},
\begin{equation}
	K = \frac{c_h}{M_P}z^\dagger H_u H_d + {\rm h.c.},
	\label{mu}
\end{equation}
as $\mu = \sqrt{3}c_h m_{3/2}$ and $B = m_{3/2}$.
Thus the framework naturally solves the $\mu/B\mu$ problem.
If one takes the gravitino mass to be as large as 10\,TeV, the SUSY flavor and CP problems 
are greatly relaxed and the cosmological gravitino problem is also ameliorated.
It also explains the 125\,GeV Higgs boson without tuning the 
$A$-parameter~\cite{Okada:1990gg,Giudice:2011cg,Ibe:2011aa,Feng:2011aa}.
Therefore the $\mathcal O(10)$\,TeV SUSY is plausible from these phenomenological point of view.

However, the model suffers from the cosmological Polonyi problem, which inevitably arises in the gravity-mediation scenario.
The Polonyi abundance is estimated as
\begin{equation}
	\frac{\rho_z}{s} \simeq \frac{1}{8}T_{\rm R}\left( \frac{z_i}{M_P} \right)^2,
	\label{rhoz}
\end{equation}
where $z_i$ is the initial amplitude, which is in general of the order of $M_P$. The reheating
temperature $T_{\rm R}$ is defined by
\begin{equation}
T_{\rm R}\;\equiv\; \lrfp{10}{\pi^2g_*}{1/4}\sqrt{\Gamma_{\rm tot} M_P},
\end{equation}
where $\Gamma_{\rm tot}$ is the inflaton decay rate, and $g_*$ counts the relativistic degrees
of freedom at the reheating. Here we have assumed that the potential for $z$ can be well 
approximated by a quadratic term for $|z| \lesssim z_i$, and that the $z$ starts to oscillate before
the reheating.
The Polonyi abundance (\ref{rhoz}) is so large that the $z$
dominates the energy density of the Universe soon after the reheating, and causes
cosmological problems.

The Polonyi decays into gauge bosons through the interaction (\ref{KW}) with the decay rate given by
\begin{equation}
	\Gamma(z \to gg) \simeq \frac{3 c_g^2}{32\pi} \frac{m_z^3}{M_P^2},
\end{equation}
where $m_z$ is the Polonyi mass at the zero temperature. 
The decay into gauginos is suppressed by $(m_{\tilde g}/m_z)^2$ or $(m_{3/2}/m_z)^2$,
where $m_{\tilde g}$ denotes the gaugino mass, and as we will see later, 
the Polonyi mass is considered to be slightly enhanced compared to $m_{\tilde g}$ or $m_{3/2}$.
We parametrize it as $m_z = c_z m_{3/2}$ with $c_z \gtrsim 1$.

The interaction (\ref{mu}) induces the decay of the Polonyi into the Higgs boson pair~\cite{Endo:2006ix},
\begin{equation}
	\Gamma(z \to HH) \simeq \frac{c_h^2}{8\pi} \frac{m_z^3}{M_P^2},
\end{equation}
while the decay into a higgsino pair is suppressed by a factor of $(m_{3/2}/m_z)^2$. 
The Polonyi also decays into a pair of gravitinos if kinematically allowed~\cite{Endo:2006zj}. The decay rate is given by
\begin{equation}
	\Gamma(z \to \psi_{3/2}\psi_{3/2}) \simeq \frac{1}{96\pi} \frac{m_z^5}{m_{3/2}^2M_P^2}.
\end{equation}

For example, if the decay into gauge bosons is the dominant decay mode, the lifetime of the Polonyi is given by
\begin{equation}
	\tau_z \simeq 1.3\times 10^{-1}c_g^{-2}\left( \frac{100\,{\rm TeV}}{m_z} \right)^3 \,{\rm sec}.
\end{equation}
If the decay into the gravitino pair is dominant, the lifetime is given by
\begin{equation}
	\tau_z \simeq 1.2\times 10^{-2}\left( \frac{100\,{\rm TeV}}{m_z} \right)^5
	\left( \frac{m_{3/2}}{10\,{\rm TeV}} \right)^2 \,{\rm sec}.
\end{equation}
The lifetime must be (much) shorter than 1\,sec in order not to spoil the success of BBN~\cite{Kawasaki:2004qu}.
Even if it decays before BBN, it dilutes the pre-existing baryon asymmetry of the Universe.
The dilution factor is roughly given by $\sim T_d / T_{\rm R}$, where $T_d$ is the Polonyi decay temperature.
The dilution factor is so large that  thermal and non-thermal leptogenesis scenarios do not work.
Therefore some involved mechanisms to create the baryon asymmetry is required if
the Polonyi problem is solved by increasing the Polonyi mass.
In the next section we will consider another attractive solution to the Polonyi problem in which there is no
late-time entropy production.

\begin{figure}
\begin{center}
\includegraphics[width=0.6\linewidth]{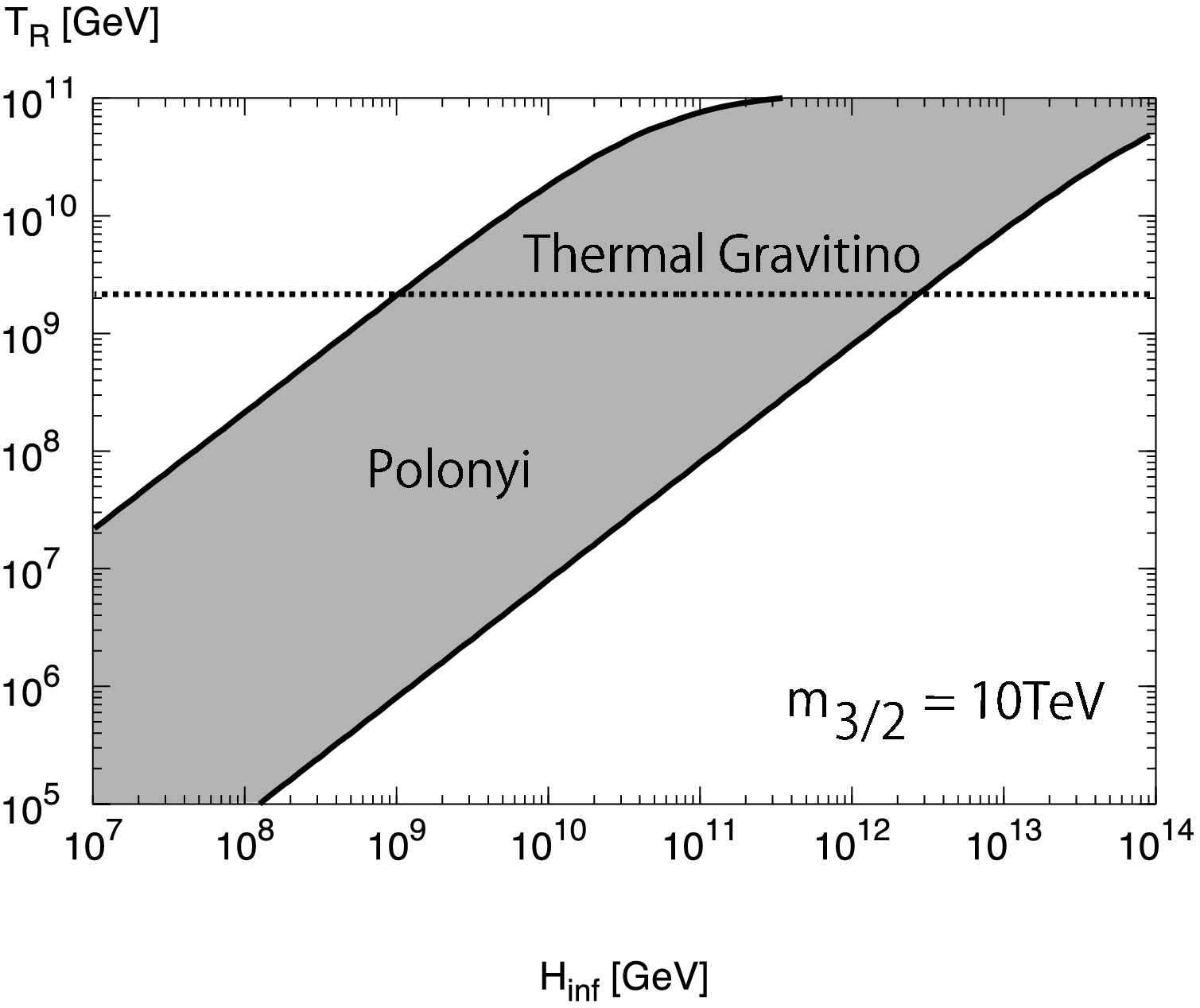}
\includegraphics[width=0.6\linewidth]{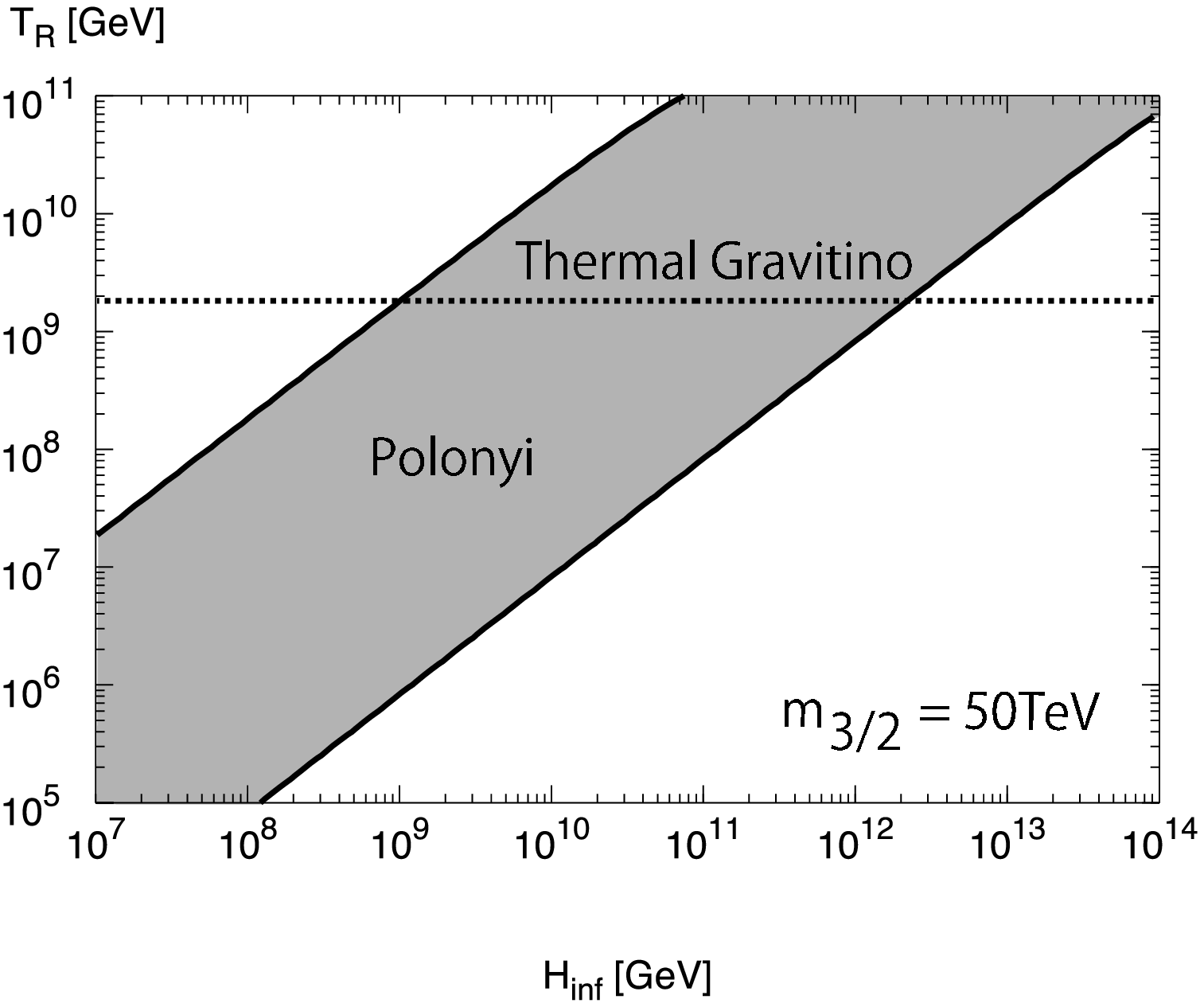}
\caption{
	Constraints on the $H_{\rm inf}$ and $T_{\rm R}$ plane from thermally produced gravitinos and
	the Polonyi coherent oscillation for $m_{3/2}=$10\,TeV (upper panel) 
	and $50$\,TeV (lower panel) with $c_g=0.1$. We set $c_h = 1$.
	The dotted line shows the upper bound from the gravitino thermal production.
	The gray band shows the upper bound on the reheating temperature from the Polonyi
	coherent oscillation and thermal production.
	The upper edge of the band corresponds to $c_z =50, c_X = 100, \Delta z = 0.1M_P/c_X$
	and lower one to $c_z =5, c_X = 50, \Delta z = M_P/c_X$.
	In the lower panel, the constraint comes from the LSP overproduction, hence 
	all the constraints disappear if the R-parity is broken by a small amount.
}
\label{fig:1}
\end{center}
\end{figure}

\section{Solution to the Polonyi problem and implications}

Now we revisit the Polonyi model in light of the recent developments in the
suppression mechanism for the moduli abundance~\cite{Nakayama:2011wqa}.

Let us introduce the inflaton fields $X$ and $\phi$, which have  $R$-charges of $+2$ and $0$, respectively.
The inflaton superpotential has the form
\begin{equation}
	W = X f(\phi),
\end{equation}
where $f(\phi)$ is some function of $\phi$. 
The $F$-term of $X$ dominates the potential energy during inflation.
Many known inflation models in supergravity fall into this category.
The Polonyi field in general couples to the inflaton fields as
\begin{equation}
	K = - c_X^2\frac{|X|^2|z-z_X|^2}{M_P^2} - c_\phi^2\frac{|\phi|^2|z-z_\phi|^2}{M_P^2},
\end{equation}
where $c_X$ and $c_\phi$ are taken to be real and positive.
The adiabatic suppression mechanism works if $c_X \gg 1$~\cite{Takahashi:2010uw}.
However, the inflaton dynamics just after the inflation induces a non-negligible amount
of the coherent oscillations of the Polonyi field, which
 is estimated as~\cite{Nakayama:2011wqa}
\begin{equation}
	\frac{\rho_z}{s} \simeq \frac{1}{8}T_{\rm R}\left( \frac{\Delta z}{M_P} \right)^2
	\left( \frac{c_\phi^4 m_z}{c_X^3 H_{\rm inf}} \right),	
	\label{CO}
\end{equation}
where $\Delta z = |z_X-z_\phi|$ and $H_{\rm inf}$ is the Hubble scale at the end of inflation.
This expression is valid for $c_\phi \gtrsim 1$.
For $c_\phi \ll 1$, there remains a contribution like (\ref{CO}) with $c_\phi$ replaced by $\mathcal O(1)$.
This is much smaller than the naive estimate (\ref{rhoz}) if $H_{\rm inf}\gg m_z$,
which is satisfied for the most known inflation models.
From this expression, we can see  that the Polonyi abundance is suppressed
for $c_X/c_\phi \gg 1$ and large inflation scale.
Hereafter we take $c_\phi=1$ for simplicity.

Now let us see how the present model is constrained from cosmological arguments.
First, gravitinos are effectively produced at the reheating, and its abundance is proportional
to the reheating	 temperature. 
If the gravitino is heavier than  the lightest SUSY particle (LSP), it is unstable and decays
emitting energetic particles.  Such late gravitino decay changes the Helium-4 abundance~\cite{Kawasaki:2004qu},
and produces LSPs non-thermally.
The Polonyi causes similar effects: the Polonyi decay may alter the standard BBN results and yield too many LSPs.
If the Polonyi decays into the gravitino, the subsequent gravitino decay also causes similar effects.
Notice that the Polonyi abundance is given by the sum of the coherent oscillation (\ref{CO})
and thermal production, the latter of which is comparable to the abundance of the (transverse components of) gravitino 
if $c_g \sim 1$.

Fig.~\ref{fig:1} shows constraints on the $H_{\rm inf}$ and $T_{\rm R}$ plane from thermally produced gravitinos and
the Polonyi coherent oscillations  and thermal production for $m_{3/2}=$10\,TeV and $c_g=0.1$ (upper panel) 
and $50$\,TeV and $c_g=0.05$ (lower panel).
The choice of relatively small $c_g$ is motivated by the existence of the focus-point region,
and the Polonyi mainly decays into gravitinos in this case. 
The dotted line shows the upper bound on $T_{\rm R}$ from the  thermal production of gravitinos.
The gray band shows the upper bound on $T_{\rm R}$ from the Polonyi
coherent oscillation and thermal production, and the width of the band represents uncertainty of the Polonyi abundance
and couplings. 
The upper edge of the band corresponds to $c_z =30, c_X = 100, \Delta z = 0.1M_P/c_X$, while
the lower one to $c_z =5, c_X = 50, \Delta z = M_P/c_X$,
where $m_z = c_z m_{3/2}$.
The Polonyi mass is varied because it is strongly coupled with the inflaton $X$ $(c_X \gg 1)$, 
and the Polonyi self interaction of the form $K \sim -c_z^2|z|^4/M_P^2$ with $c_z \gg 1$ is expected
in the K\"ahler potential.
With the present parameter choice, the Bino is the LSP of mass  $360$\,GeV (upper panel) and $900$\,GeV (lower panel).
The thermal relic abundance of the Bino LSP is not taken into account in Fig.~\ref{fig:1}, because it strongly depends on
the mass spectrum. For instance, 
if it has a sizable mixing with higgsino or wino, the thermal relic abundance can be smaller than the present DM abundance.
(In the latter case, we need to relax the GUT relation on the gaugino mass.) We note that,
in the lower panel, the constraint comes from the LSP overproduction from the gravitino/Polonyi decay, hence 
all the constraints disappear if the R-parity is broken by a small amount.

It is seen that the reheating temperature of $T_{\rm R}\simeq 10^9$\,GeV is allowed
for $H_{\rm inf} \gtrsim 10^9$--$10^{12}$\,GeV.
It is important that we do not need any additional late-time entropy production for solving the Polonyi problem.
Thus the conventional Polonyi model for the gravity-mediation for relatively heavy SUSY scale
of $\mathcal O(10)$\,TeV can be compatible with leptogenesis scenario
once we assume the Polonyi coupling to the inflaton is enhanced.

\section{Inflation model}

Now let us see if the above solution works in some known inflation models in supergravity.
In particular, we will show that there are consistent parameter regions where
thermal~\cite{Fukugita:1986hr} or non-thermal~\cite{Lazarides:1991wu,Kumekawa:1994gx,Asaka:1999yd} leptogenesis scenario works, 
avoiding the Polonyi and gravitino problems.

\subsection{Hybrid inflation }

\begin{figure}
\begin{center}
\includegraphics[width=0.6\linewidth]{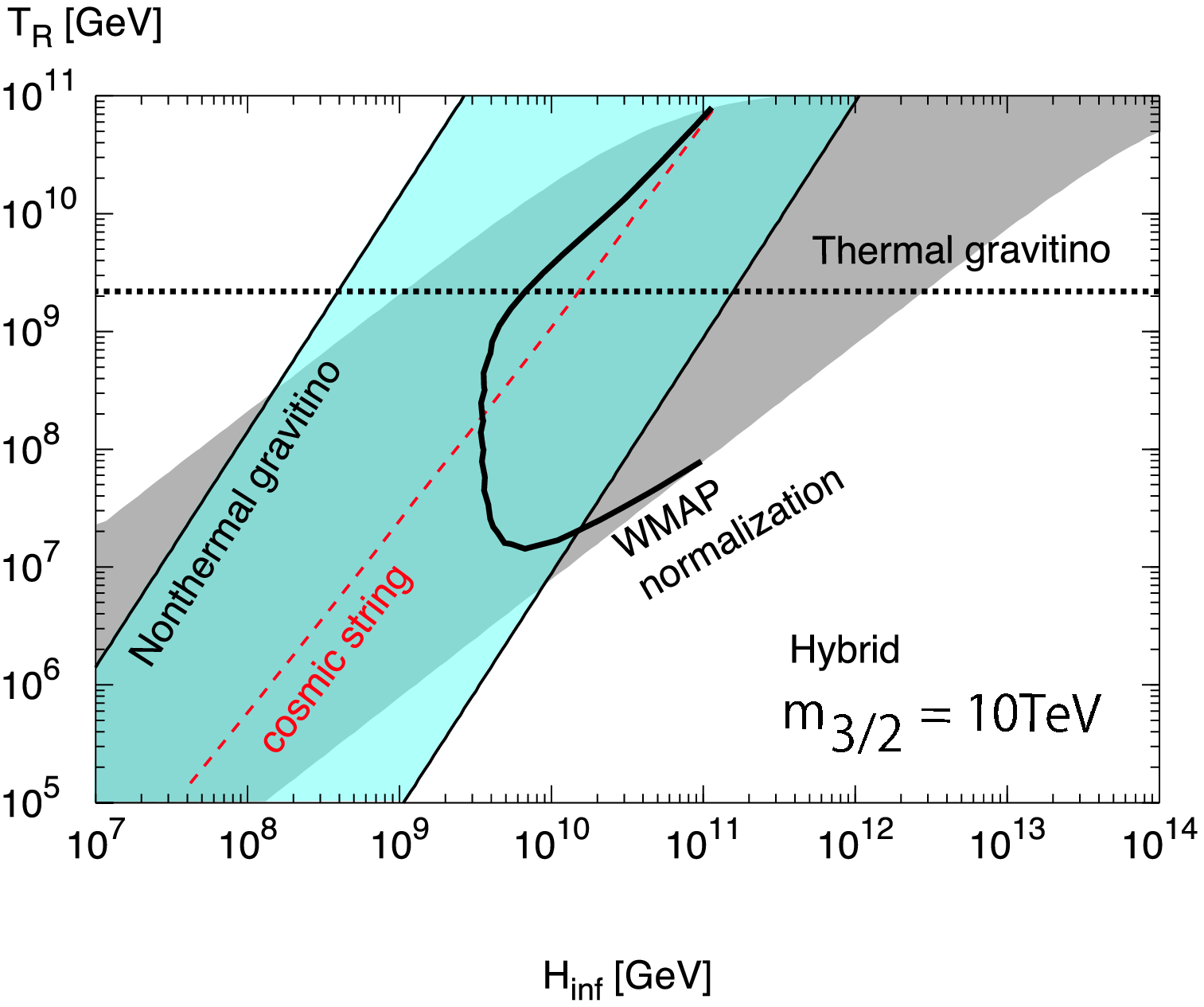}
\includegraphics[width=.6\linewidth]{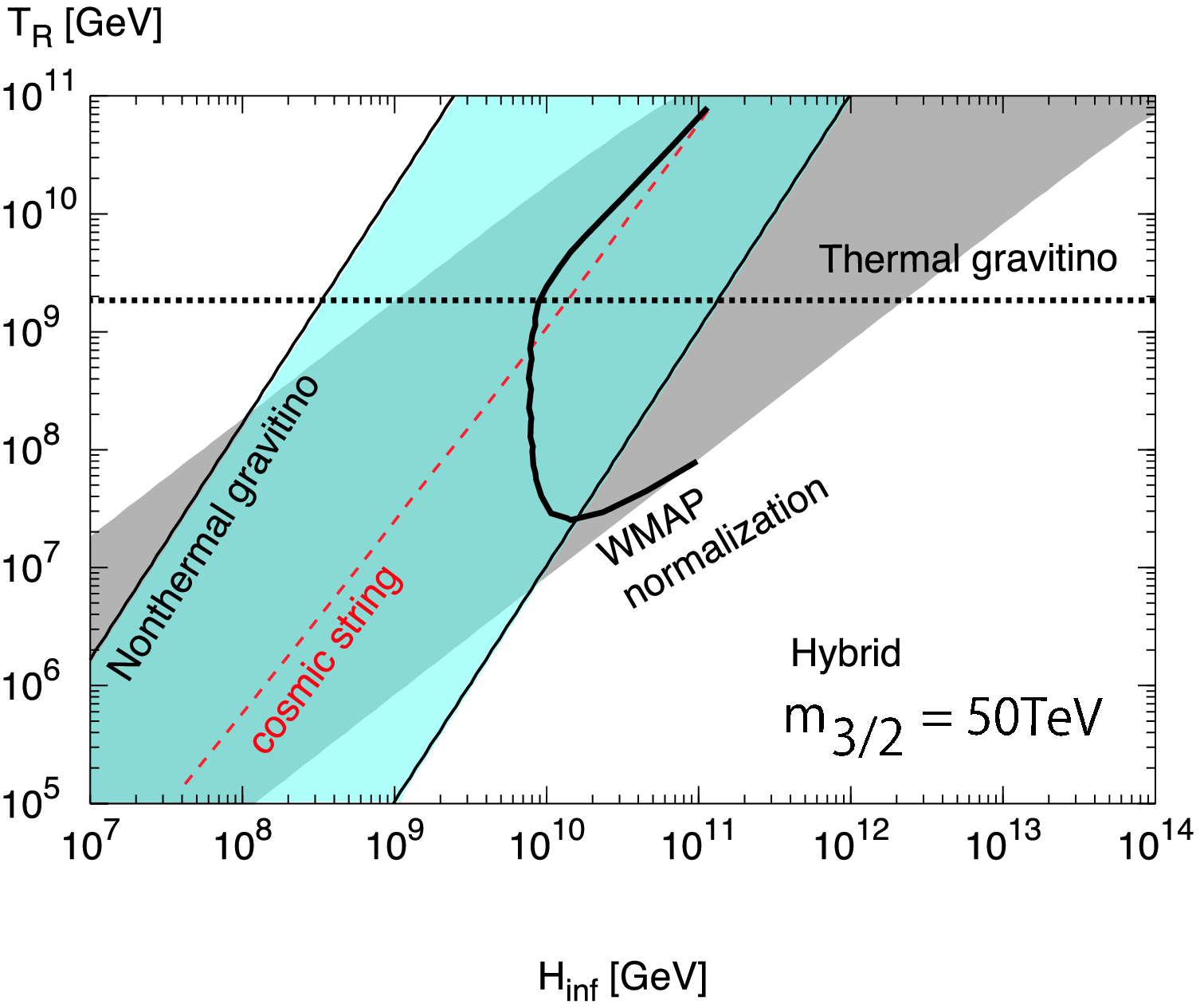}
\caption{
	Constraints in the $H_{\rm inf}$ and $T_{\rm R}$ plane for the hybrid inflation model
	with $m_{3/2}=$10\,TeV (upper panel) and $50$\,TeV (lower panel).
	The red dashed line shows the lower bound on $T_{\rm R}$ from the cosmic string.
The blue band shows the lower bound on $T_{\rm R}$ 
from the non-thermal gravitinos for $c_{\phi zz}=1$ (upper edge) and $0.1$ (lower edge).
The meanings of gray band and the black dotted line are same as Fig.~\ref{fig:1} :
they set upper bounds on $T_{\rm R}$ from the Polonyi and thermal gravitino.
The density perturbation with a correct magnitude is generated on the solid line
denoted by ``WMAP normalization".
}
\label{fig:hybrid}
\end{center}
\end{figure}

First, let us consider the SUSY hybrid inflation model~\cite{Copeland:1994vg,Dvali:1994ms,Linde:1997sj}.
The superpotential is given by
\begin{equation}
	W = \kappa X (\phi\bar\phi - M^2) + W_0,
\end{equation}
where $W_0=m_{3/2}M_P^2$.
The waterfall fields, $\phi$ and $\bar\phi$, can be identified with the Higgs fields which break
U(1)$_{\rm B-L}$ gauge symmetry.
This model, including the constant term $W_0$, was analyzed in detail in Refs.~\cite{Buchmuller:2000zm,Senoguz:2004vu,Nakayama:2010xf}.
We assume that the inflaton dominantly decays into the right-handed neutrinos $N_i$ $(i=1,2,3)$ through the interaction
\begin{equation}
	W = \frac{1}{2}y_{i} \phi N_i N_i.
	\label{yNN}
\end{equation}
The inflaton decay rate into the right-handed (s)neutrino pair is given by
\begin{equation}
	\Gamma (\phi \to N_1N_1,\tilde N_1\tilde N_1) \simeq \frac{1}{64\pi}y_{1}^2m_\phi,
	\label{Gamma}
\end{equation}
where we have taken into account a mixing between $X$ and $\phi$ (and $\bar \phi$) due to the
constant term~\cite{Kawasaki:2006gs}.
Here we consider only the decay into the lightest right-handed neutrino.
On the other hand, the inflaton decays into a pair of gravitinos through the interaction in the 
K\"ahler potential~\cite{Kawasaki:2006gs,Dine:2006ii,Endo:2006tf},
\begin{equation}
	K = \frac{1}{2M_P^2}(c_{\phi zz}^2 |\phi|^2 +c_{\bar\phi zz}^2 |\bar\phi|^2) zz + {\rm h.c.}
\end{equation}
The decay rate into the gravitino pair is given by~\cite{Endo:2006tf}
\begin{equation}
	\Gamma_{\rm grav}\equiv\Gamma(\phi \to \psi_{3/2}\psi_{3/2}) = \frac{1}{32\pi} \lrfp{c_{\phi zz}^2 + c_{\bar \phi zz}^2}{2}{2}
	\left( \frac{\langle\phi\rangle}{M_P} \right)^2\frac{m_\phi^3}{M_P^2},
\end{equation}
where the mixing between $X$ and $\phi$ (and $\bar \phi$) is taken into account~\cite{Kawasaki:2006gs}.
Notice that the same interaction induces the inflaton decay into the Polonyi pair $(\phi\to zz)$ with the same decay rate.
Since each Polonyi field mainly decays into a pair of the gravitino, 
the gravitino abundance produced non-thermally by the inflaton decay is given by
\begin{equation}
	Y_{3/2} = \frac{3}{2}\frac{T_{\rm R}}{m_\phi} \frac{3\Gamma_{\rm grav}}{\Gamma_{\rm tot}},
	\label{Ygrav}
\end{equation}
where the total decay rate is approximately given by $\Gamma_{\rm tot} \approx \Gamma (\phi \to NN)$.
This imposes severe constraints on the parameter space.
We have scanned parameters $(\kappa, M)$, which are rewritten in terms of
$H_{\rm inf}$ and $T_{\rm R}$ through the relation $H_{\rm inf}=\kappa M^2 /\sqrt{3}M_P$
and $T_{\rm R} = (10/\pi^2 g_*)^{1/4}\sqrt{\Gamma_{\rm tot}M_P}$.
We have also fixed $m_N=0.02m_\phi$: the non-thermal leptogenesis works for
$T_{\rm R}\gtrsim 10^8$\,GeV in this case.
Fig.~\ref{fig:hybrid} shows constraints in the $H_{\rm inf}$ and $T_{\rm R}$ plane for the hybrid inflation model
with $m_{3/2}=$10\,TeV (upper panel) and $50$\,TeV (lower panel).
The red dashed line shows the lower bound on $T_{\rm R}$ from the cosmic string.
The blue band shows the lower bound on $T_{\rm R}$ 
from the non-thermal gravitinos for $c_{\phi zz}=1$ (upper edge) and $0.1$ (lower edge).
The meanings of the gray band and the black dotted line are same as Fig.~\ref{fig:1} :
they set upper bounds on $T_{\rm R}$ from the Polonyi and thermal gravitino.
The density perturbation with a correct magnitude is generated on the solid line labels by ``WMAP normalization".

It is seen that there is a consistent parameter regions
around $H_{\rm inf}\sim 5\times 10^9$\,GeV and $T_{\rm R}\sim 10^9$\,GeV
where the Polonyi problem is solved within the framework of SUSY hybrid inflation model.
Note that the constraints from the Polonyi and gravitinos disappear if the R-parity is broken slightly for $m_{3/2} \gtrsim 30$\,TeV,
as already explained.

\begin{figure}
\begin{center}
\includegraphics[width=0.6\linewidth]{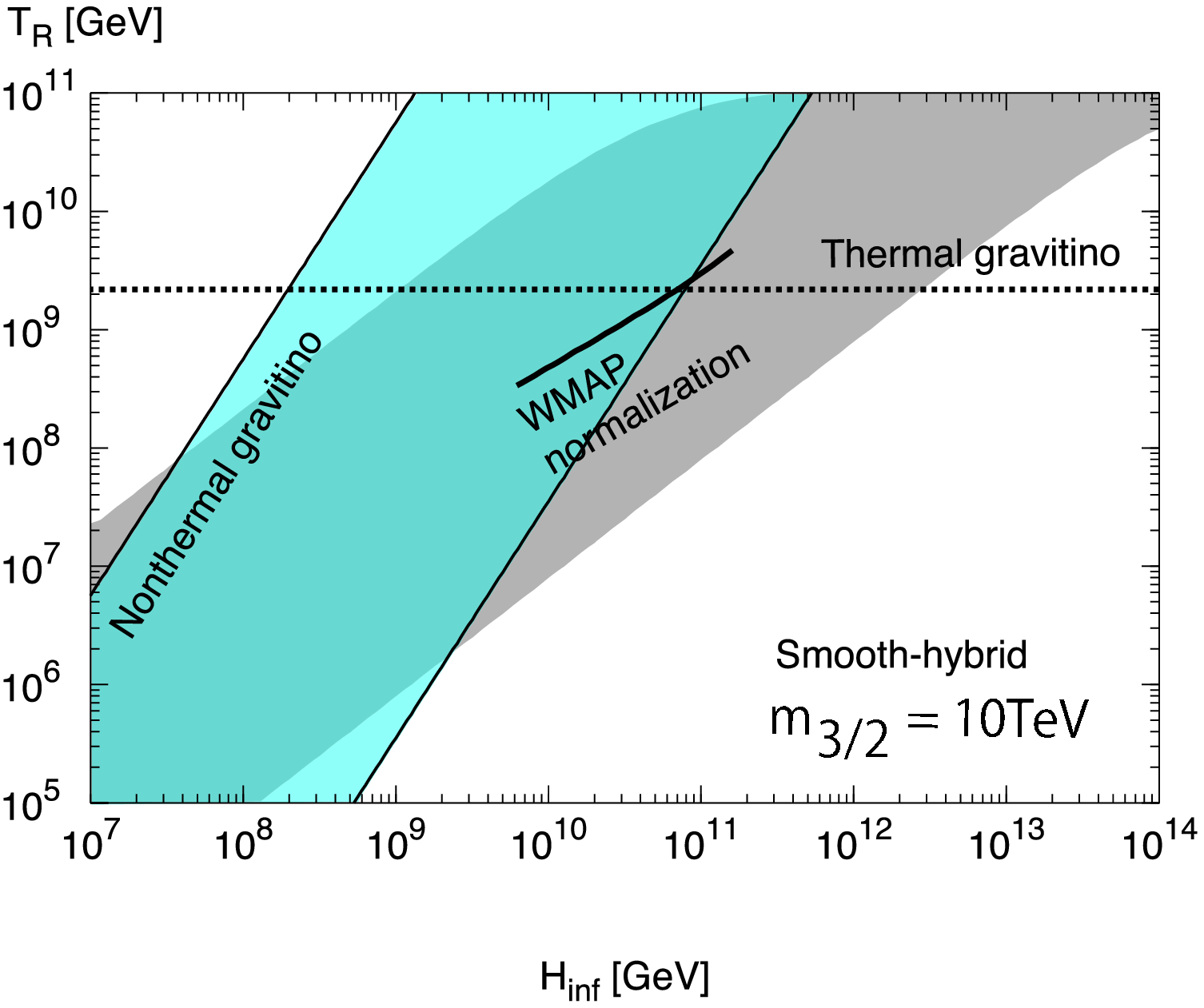}
\includegraphics[width=0.6\linewidth]{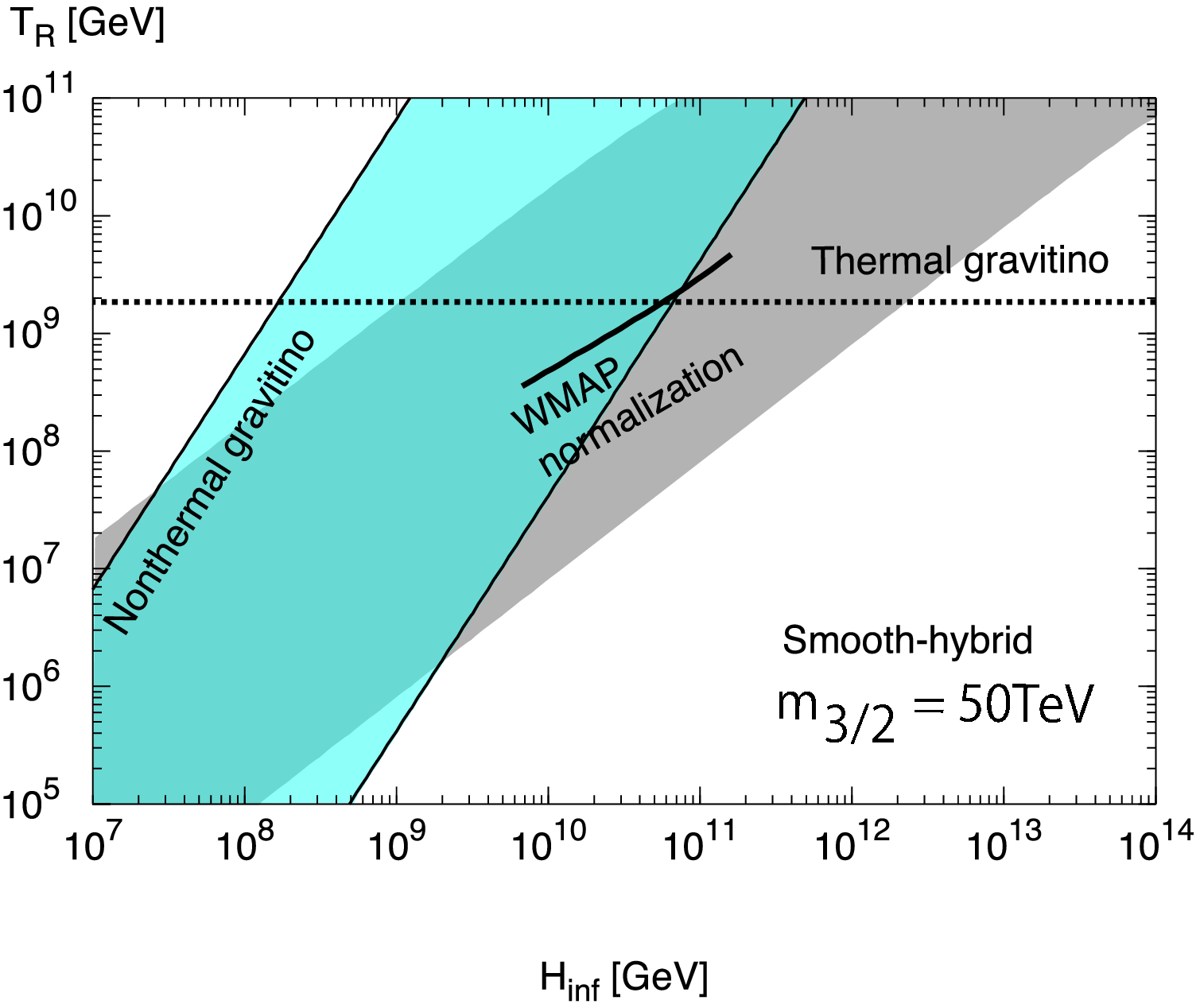}
\caption{
	Same as Fig.~\ref{fig:hybrid}, but for the smooth-hybrid inflation model.
	The blue band shows the lower bound on $T_{\rm R}$ 
from the non-thermal gravitinos for $c_{\phi zz}=1$ (upper edge) and $0.05$ (lower edge).
	}
\label{fig:smooth}
\end{center}
\end{figure}

\subsection{Smooth hybrid inflation }

Let us consider the smooth-hybrid inflation model~\cite{Lazarides:1995vr} where
the inflaton superpotential is given by
\begin{equation}
	W = X\left( \mu^2 - \frac{(\phi\bar\phi)^m}{M^{2m-2}} \right) + W_0,  \label{smooth}
\end{equation}
where $m \geq 2$ is an integer.
The model has a discrete symmetry $Z_m$ under which $\phi \bar\phi$ has a charge $+1$ and $X$ has a zero charge.
This model has an advantage that it does not suffer from problematic topological defects formation
since the $\phi$ and $\bar\phi$ have nonzero VEVs during inflation and topological defects are inflated away.
Hereafter we consider the case of $m=2$ for simplicity.
Results do not much affected by this choice.
The gravitino abundance is similarly estimated by Eq.~(\ref{Ygrav}).

The inflaton can decay into ordinary particles through non-renormalizable interactions, for example,
\begin{equation}
	K=\frac{|\phi|^2 H_u H_d}{M_c^2}+{\rm h.c.},
\end{equation}
with cutoff parameter $M_c$.
The decay rate into the Higgs boson and higgsino pair is given by
\begin{equation}
	\Gamma(\phi \to HH) = \frac{1}{16\pi}
	\left( \frac{\langle\phi\rangle}{M_c} \right)^2\frac{m_\phi^3}{M_c^2},
\end{equation}
where the mixing between $X$ and $\phi$ (and $\bar \phi$) is taken into account. 
If the right-handed neutrino mass is not much smaller than the inflaton mass, 
the decay rate into them through the operator $K=|\phi|^2|N|^2/M_c^2$ is comparable to the above expression.

We have scanned the parameters $(\mu, M)$, in the range $\mu < M$ 
so that the effective theory (\ref{smooth}) below the scale $M$ remains valid,
which are rewritten in terms of $H_{\rm inf}$ and $T_{\rm R}$ through the relation $H_{\rm inf}=\mu^2 /\sqrt{3}M_P$
and $T_{\rm R} = (10/\pi^2 g_*)^{1/4}\sqrt{\Gamma_{\rm tot}M_P}$.
We have fixed $M_c=6\times 10^{17}$\,GeV.
Fig.~\ref{fig:smooth} shows constraints on the $H_{\rm inf}$ and $T_{\rm R}$ plane for the hybrid inflation model
with $m_{3/2}=$10\,TeV (upper panel) and $50$\,TeV (lower panel).
The blue band shows the lower bound on $T_{\rm R}$ 
from the non-thermal gravitinos for $c_{\phi zz}=1$ (upper edge) and $0.05$ (lower edge).
The meanings of the gray band and the black dotted line are same as Fig.~\ref{fig:1} :
they set upper bounds on $T_{\rm R}$ from the Polonyi and thermal gravitino.
The WMAP normalization for the density perturbation is satisfied on the solid line.
The scalar spectral index $n_s$ also fits well with the WMAP result: $n_s \sim 0.968 \pm 0.012$~\cite{Komatsu:2010fb}.

It is seen that there is a consistent parameter regions
around $H_{\rm inf}\sim 10^{10-11}$\,GeV and $T_{\rm R}\sim 10^{8-9}$\,GeV
where the Polonyi problem and the gravitino problem are solved within the framework of smooth-hybrid inflation model.
Note again that the constraints from the Polonyi and gravitinos disappear if R-parity is broken slightly for $m_{3/2} \gtrsim 30$\,TeV,
as already explained.

\section{Conclusion}

We have revisited the Polonyi model for gravity mediation with a relatively high-scale SUSY breaking
of $\mathcal O(10)$\,TeV.
The Higgs boson mass of around 125\,GeV indicated by the recent LHC data is naturally explained
in this framework, while constraints from flavor/CP violating processes are alleviated.
The model, however, is plagued with the notorious cosmological Polonyi problem.
We have shown that the Polonyi problem is solved once we assume the relatively enhanced coupling
of the Polonyi to the inflaton. We have also considered explicit inflation models (hybrid and smooth hybrid
inflation), and shown that there is a parameter space where thermal and/or non-thermal leptogenesis scenarios work successfully,
avoiding the Polonyi and gravitino problems. 
Thus, our solution revives the conventional Polonyi model as a realistic SUSY breaking model.

\begin{acknowledgments}
This work was supported by the Grant-in-Aid for Scientific Research on
Innovative Areas (No. 21111006) [KN and FT], Scientific Research (A)
(No. 22244030 [KN and FT], 21244033 [FT], 22244021 [TTY]), and JSPS Grant-in-Aid for
Young Scientists (B) (No. 21740160) [FT].  This work was also
supported by World Premier International Center Initiative (WPI
Program), MEXT, Japan.

\end{acknowledgments}

   
 
 \end{document}